\documentclass[preprint,showpacs,aps,superscriptaddress,nofootinbib]{revtex4}
\usepackage{graphicx}
\begin{document}
\setlength{\topmargin}{-2ex}
 
 \title{The accelerated expansion of the Universe as a quantum cosmological effect }
\author{N. Pinto-Neto}
 \affiliation{Centro Brasileiro de Pesquisas F\'{\i}sicas, \\
Laborat\'orio de Cosmologia e F\'{\i}sica Experimental de Altas Energ\'{\i}as, \\
Rua Dr.\ Xavier Sigaud 150, Urca 22290-180 -- Rio de Janeiro, RJ -- Brasil}
\email{nelsonpn@cbpf.br}
\author{E. Sergio Santini}
\affiliation{Universidade Federal do Esp\'{\i}rito Santo, \\ 
Departamento de F\'{\i}sica, Centro de Ci\^encias Exatas,  \\ 
Av. Fernando Ferrari, s/n, Campus de Goiabeiras
 29060-900, Vit\'oria, ES -- Brasil}

\affiliation{Centro Brasileiro de Pesquisas F\'{\i}sicas, \\
     Laborat\'orio de Cosmologia e F\'{\i}sica Experimental de Altas Energ\'{\i}as, \\
    Rua Dr.\ Xavier Sigaud 150, Urca 22290-180 -- Rio de Janeiro, RJ -- Brasil}
\affiliation{ Comiss\~ao Nacional de Energia Nuclear \\ Diviss\~ao de Instala\c c\~oes Radiativas \\ Rua General Severiano 90, Botafogo 22290-901 -- Rio de Janeiro, RJ -- Brasil}
\email{santini@cbpf.br} 
    \date{\today}

    \begin{abstract}
    We study the quantized Friedmann-Lema\^{\i}tre-Robertson-Walker (FLRW)
    model  minimally coupled to a free massless scalar field.
    In a previous paper, \cite{fab2}, solutions of this model were constructed as 
    gaussian superpositions of negative and positive modes solutions of the 
Wheeler-DeWitt equation, and quantum bohmian trajectories were obtained in the 
framework of the Bohm-de Broglie (BdB) interpretation of quantum cosmology.
    In the present work, we analyze the quantum bohmian trajectories of a
    different class of gaussian packets. We are able to show that this new class 
generates bohmian trajectories which begin classical (with decelerated expansion),
    undergo an accelerated expansion in the middle of its evolution due to the presence
    of quantum cosmological effects in this period, and return to its classical
    decelerated expansion in the far future. We also show that the relation between 
luminosity distance and redshift in the quantum cosmological model can be made
    close to the corresponding relation coming from the classical model suplemented
    by a cosmological constant, for $z<1$. These results suggest the posibility of 
interpreting the present observations of high redshift supernovae as the 
manifestation of a quantum cosmological effect.
    \end{abstract}
    \maketitle


    \section{Introduction}
    Recent measurements of high redshift supernovae \cite{SN1}\cite{SN} indicate
    that
    the Universe is presently in accelerated expansion and
    not in a decelerated one, as
    was firmly believed by
    cosmologists since Hubble's
    observation and its interpretation in the framework of
    General Relativity (GR).
    This was a spectacular break through in our understanding (or misunderstanding)
    of the
    Universe, and it became
    a major task to Cosmology and  Astrophysics to explain this unexpected fact.
    Staying in the domain of classical GR and, consequently, considering the
    Friedmann's equations as valid, the only way
    to explain such present acceleration of the Universe is by considering the
    existence of some
    negative pressure dark energy
    \cite{SN1}\cite{triangle}\cite{garnavich}\cite{wang}.This cosmic
    dark energy opposes the self-atraction of matter and is
    causing the expansion of the universe to be positively accelerated\cite{triangle}.
    The most obvious candidate to be such dark energy is the cosmological constant 
    and/or
    the vacuum quantum fluctuations of fields, which do have negative pressure.
    However, theorists estimates that the zero point
    energies of the quantum fields must be at least 55
    orders of magnitude larger than the critical density value. Hence,
    there must exist some yet unknown profound
    theoretical reason for the many contributions to the effective value of
    the cosmological constant be cancelled out to yield a number 55 orders of
    magnitude less then expected, or even zero.
    This is known as the
    cosmological constant problem. Some theorists believe that some profound
    symmetry requirements can be found to explain an exact cancellation, but not
    a partial one with an extreme fine tuning.    
    In the case where the effective cosmological constant is indeed exactly zero,
    there were proposed some candidates in order to
    explain the present accelerated expansion of the Universe as, for example,
    a very light,
    evolving scalar field called quintescence \cite{quint}\cite{jer}.

    Another way to tackle this problem is by considering that presently, at
    cosmological scales,
    classical GR is not valid. In other words, instead of changing the
    right-hand-side (RHS) of Einstein's equations
     by introducing some new negative pressure fluid, one could try to find physical reasons 
     which justify the modification of its left-hand-side (LHS) accordingly.
     How this can be done?

    In early works \cite{fab2}\cite{fab}, a quantum
    minisuperspace model containing a free massless scalar field
    minimally coupled to gravity in a
    FLRW geometry was studied. These models
    were interpreted in the framework of the ontological Bohm-de Broglie
    (BdB) interpretation of quantum mechanics,
    \cite{bohm1}\cite{bohm2}\cite{hol}, in order to extract predictions
    from the wave function of the Universe.
    This interpretation avoids many conceptual difficulties inherent to the
    application of the Copenhagen interpretation to the quantization of the
    whole Universe, where no place for a classical domain exists. The
    BdB interpretation does not need a classical domain outside the quantized
    system to generate the physical facts out of potentialities
    (the facts are there {\it ab initio}), and hence it can be
    applied to the Universe as a whole \footnote{Other alternative
    interpretations can be used in quantum cosmology, as the
    many worlds interpretation of quantum mechanics \cite{eve}}.
    The solutions of the Wheeler-DeWitt equation for
    such scalar tensor model contain positive and negative
    frequency modes, the first leading to an expanding universe, and the second
    to a contracting one.
    There were constructed some particular superpositions
   mixing negative and positive frequency modes.
    In Ref. \cite{fab2}, gaussian superpositions were studied and, for the
    case of flat
    spatial section, the Bohm guidance equations  were reduced to a
    dynamical system. In this way the quantum trajectories were studied, emerging
    the following three kind of scenarios: periodic
    solutions, representing oscilating universes, bouncing universes, and models
    with a big bang followed by a big crunch.
    The bouncing universes contract classicaly from infinity until a minimum size,
    where quantum effects become important acting as a
    repulsive force avoiding the singularity, expanding afterwards to an infinite
    size, approaching the classical expansion
    as long as the scale factor increases.  For the periodic
    solutions, the quantum effects are always important, and they do not grow
    enough to yield a large Universe as ours. The models with a big bang followed
    by a big crunch behave as the
    classical solutions for small values of the scale factor, but display quantum
    behaviour for large scale factor. These quantum effects are responsible
    for the turning over of these solutions from decelerated expansion to
    contraction. Near the big crunch, the quantum effects are again negligible.
    Bohmian trajectories which behave classically for small scale factors
    but quantically for large scale factors where already found in Ref. \cite{fab}.
    This is not surprising as it is well known \cite{hartle} that a large universe
    behaves classically or quantically depending on its initial quantum state.
    After these remarks, the natural question one can ask is if it is possible
    that quantum
    cosmological effects at large scales can mimic a negative pressure fluid and
    yield a positive acceleration for the whole Universe. The aim of this paper
    is to show with a simple model that it is
    indeed possible for some suitable initial quantum states of the universe.
    We take the flat model considered in Ref. \cite{fab2},
    and we consider another gaussian
    superposition of negative and positive modes solutions
    of the Wheeler-DeWitt equation. We write
    the Bohm guidance equations, which are reduced to a
    dynamical system, and we analyze the bohmian trajectories in configuration
    space. We find the two following scenarios,
    depending on the initial conditions: oscillating universes without
    singularities and with relative small amplitudes of oscillation, and
    universes which arise classically
    from a singularity, experience quantum effects
    in the middle of its expansion, and recover  its classical behaviour for
    large
    values of the scale factor. We concentrate our attention on
    these solutions and we study the epoch where
    the quantum effects are important.
    We calculate its acceleration and  explore its behaviour as a function of the
    scalar field $\phi$ and of the logarithm of the scale factor,
    $\alpha\equiv \ln(a)$.
    We find that a positive acceleration of the universe
    can be obtained in such models, whose value can be adjusted by the choice of the free parameters of the model.
    This positive
    acceleration is
    a quantum effect. The mechanism is driven by the quantum potential, which
    appears in the modified quantum Einstein-Hamilton-Jacobi
    equation and modifies
    the usual classical trajectories. In this model, the acceleration is not
    forever: in the future, the universe recovers its classical deccelerated
    expansion.
    In this way, we present a possible alternative explanation for the 
    accelerated expansion of the Universe today.
    Note that this explanation
    is based on quantum effects not
    only present in
    the scalar field, as described within a different approach in
    Ref.\cite{parker}, but also in the geometry
    itself.

    The article is organized as follows. In Sec. II, we describe the classical model
    and we quantize it. In Sec. III, we introduce the Bohm-de Broglie interpretation
    of the quantized minisuperspace model presented in Sec. II. We study gaussian
    superpositions
    of the quantum solutions previously found, and we obtain the
    corresponding quantum bohmian
    trajectories. In Sec. IV
    we analyze the beahviour of the
    acceleration of the scale factor in the
    quantum bohmian
    trajectories, first qualitatively, by showing some period
    in the history of the model where the acceleration of its expansion is positive,
    then quantitatively, by comparing the curve relating the luminosity distance
    with redshift in the quantum model with the corresponding curve coming from the
    classical model
    suplemented by a cosmological constant.
    Sec. IV is for discussions and conclusions.

    \section{Classical and quantum minisuperspace models }
    In this section we make an overview of the models studied in Ref. \cite{fab2}

    \subsection{Classical Model.}
    We start from the Lagrangian

    \begin{equation}\label{2}
    L=\sqrt{-g} \biggr[R-{1\over 2}\phi_{;\rho}\phi^{;\rho}\biggl] ,
    \end{equation}
    We consider the FLRW metric given (in isotropical coordinates) by

    \begin{equation}\label{3}
    ds^2=-N^2dt^2+\frac{a(t)^2}{(1+\frac{\epsilon}{4}r^2)^2}
    \biggr[dr^2+r^2(d\theta^2+\sin ^2\theta d\phi^2)\biggl] ,
    \end{equation}
   the quantity $\epsilon$ being the spatial curvature with values $0, 1,-1$
    for flat, spherical
    and hyperbolic spatial sections, respectively. This line element will give,
    after inserting
    it in the Lagrangian (\ref{2}), the following action:

    \begin{equation}\label{4}
    S=\frac{3V}{4 \pi l^2_P}\int\frac{N a^3}{2}\biggr(\frac{-\dot{a}^2}{N^2 a^2}+
    {1\over 2}\frac{\dot{\phi}^2}{6N^2}+\frac{\epsilon}{a^2}      \biggr)dt ,
   \end{equation}
   where we have set $\hbar=c=1$ and $\dot{}\equiv \frac{d}{dt}$. The
   quantity $V$ is the volume divided by $a^3$ of the
   spacelike hypersurfaces, which are supposed to be closed, and $l_P$ is the
   Planck length.
   The total volume $V$ depends on the value of $\epsilon$ and on the topology of
   the hypersurfaces. For $\epsilon=0$, $V$ can have any value because the
   fundamental polyhedra of closed
   $\epsilon=0$ hypersurfaces
   can have arbitrary size \cite{dl}. For the case $\epsilon=1$ and topology $S^3$
   we have $V=2\pi^2$.
   Defining $\beta^2=\frac{4\pi l^2_P}{3V}$,
   $\bar{\phi}\equiv\sqrt{\phi/12}$, and omiting the bars,
   we obtain for the Hamiltonian:

   \begin{equation}\label{5}
   H=N\biggr(-\beta^2\frac{p^2_a}{2a}+\beta^2\frac{p^2_{\phi}}{2a^3}-
   \epsilon\frac{a}{2\beta^2}\biggl)
   \end{equation}
where $p_a$ and $p_\phi$ are the moments canonically conjugate to $a$ and $\phi$
   respectivelly, given by:

\begin{equation}\label{6}
   p_a=-\frac{a\dot{a}}{\beta^2N}
   \end{equation}

\begin{equation}\label{7}
  p_\phi=\frac{a^3\dot{\phi}}{\beta^2N}
 \end{equation}

A dimensionless scale factor is defined by
   $\tilde{a}\equiv\frac{a}{\beta}$ and the Hamiltonian becomes,
   omitting the tilde,

\begin{equation}\label{8}
   H=\frac{N}{2\beta}\biggr(-\frac{p^2_a}{a}+\frac{p^2_{\phi}}{a^3}-\epsilon a
   \biggl)
   \end{equation}
As $\beta$ is a multiplicative constant in the hamiltonian, we can set $\beta=1$
without
   any loss of generality, keeping in mind that the scale factor which appears in
   the
   metric is $a_{\rm phys}\equiv\beta a$, not $a$. Defining now $
   \alpha \equiv \ln(a)$, we simplify
   the Hamiltonian obtaining:

\begin{equation}\label{9}
 H=\frac{N}{2 e^{3\alpha}}\biggr[-p^2_{\alpha}+p^2_{\phi}-\epsilon e^{4\alpha} \biggl] ,
 \end{equation}
where

\begin{equation}\label{10}
 p_{\alpha}=-\frac{e^{3\alpha}\dot{\alpha}}{N} ,
 \end{equation}

   \begin{equation}\label{11}
 p_\phi=\frac{e^{3\alpha}\dot{\phi}}{N} ,
\end{equation}
This Hamiltonian does not depend explicitly on $\phi$. Hence, $p_{\phi}$ is a
constant of motion, which we will call $\bar{k}$. The classical solutions
in the gauge $N=1$ can now be listed:
   \vspace{0.4cm}
  \subsubsection{Flat model, $\epsilon=0$.}
In configuration space, the classical solutions are:

\begin{equation}\label{12}
\phi=\pm\alpha+c_1 ,
 \end{equation}
   where $c_1$ is an integration constant. In terms of cosmic time $t$ they read:

 \begin{equation}\label{13}
a=e^{\alpha}=(3\bar{k}t)^{\frac{1}{3}} ,
 \end{equation}

\begin{equation}\label{14}
 \phi=\frac{\ln(t)}{3}+c_2  .
 \end{equation}
  These are solutions forever contracting or expanding from a singularity,
 depending on the signal of $\bar{k}$, without any inflationary epoch.
   \vspace{0.4cm}

\subsubsection{Spherical model, $\epsilon=1$.}
   In this case we have,

\begin{equation}\label{15}
a=e^{\alpha}=\frac{\bar{k}}{\cosh(2\phi-c_1)} ,
\end{equation}
where $c_1$ is an integration constant. Conservation of $p_{\phi}$ implies that

\begin{equation}\label{16}
\bar{k}=e^{3\alpha}\dot{\phi} .
\end{equation}
  These solutions describe universes expanding from a singularity till
a maximum size and contracting again to a big crunch. Near the
 singularity, these solutions behave as in the flat case. There is no inflation.
\vspace{0.4cm}

\subsubsection{Hyperbolic model, $\epsilon=-1$.}

The classical solutions in configuration space are:

\begin{equation}\label{17}
a=e^{\alpha}=\frac{\bar{k}}{|\sinh(2\phi-c_1)|} ,
\end{equation}
  where $c_1$ is an integration constant. Again, from the conservation of
$p_{\phi}$ we get

   \begin{equation}\label{18}
 \bar{k}=e^{3\alpha} \dot{\phi} .
 \end{equation}
  These solutions describe universes contracting forever to, or expanding
forever from, a
singularity. Near the singularity, these solutions behave as in the flat
case. There is no inflation phase.
The cosmic time dependence is complicated in the cases {\it 2} and {\it 3}
and we will not write it here.

\subsection{Quantization.}
Let us quantize the model following the Dirac procedure \cite{dirac}.
The constraints
become conditions imposed on the possible states of the quantum
system. The operator version of the Hamiltonian (\ref{9}), obtained by setting
$\hat{\alpha}\rightarrow -\imath \frac{\partial}{\partial \alpha}$ and
$\hat{\phi}\rightarrow -\imath \frac{\partial}{\partial \phi}$, must annihilate
the wave function $\Psi$. Choosing a factor ordering which make it covariant
through
field redefinitions, the quantum constraint, i.e. the Wheeler-DeWitt equation,
reads

\begin{equation}\label{19}
-\frac{\partial^2 \Psi}{\partial \alpha^2}+
\frac{\partial^2 \Psi}{\partial \phi^2}+\epsilon e^{4\alpha}\Psi=0 ,
\end{equation}
whose general solution can be written as

\begin{equation}\label{20}
\Psi(\alpha,\phi)=\int_{-\infty}^{\infty} F(k)A_k(\alpha)B_k(\phi)dk ,
\end{equation}
where $k$ is a separation constant which in the classical limit corresponds
to $\bar{k}$, $F(k)$ is an arbitrary function of $k$, the function $B_k(\phi)$
reads

\begin{equation}\label{21}
B_k(\phi)=b_1 e^{ik\phi}+b_2 e^{-ik\phi},
\end{equation}
and, for $\epsilon=0$, the function $A_k(\alpha)$ is given by

\begin{equation}\label{22}
A_k(\alpha)=a_1 e^{ik\alpha}+a_2 e^{-ik\alpha},
\end{equation}
while for  $\epsilon=1$ it is

\begin{equation}\label{23}
A_k(\alpha)=a_1 I_{ik/2}\frac{e^{2\alpha}}{2}+a_2 K_{ik/2} \frac{e^{2\alpha}}{2},
\end{equation}
and for $\epsilon=-1$ it reads

\begin{equation}\label{24}
A_k(\alpha)=a_1 J_{ik/2}\frac{e^{2\alpha}}{2}+a_2 N_{ik/2} \frac{e^{2\alpha}}{2} .
\end{equation}
The functions $J, N, I, K $ are Bessel and modified Bessel functions of first and
second kind.

\section{The Bohm-de Broglie interpretation of the quantum model}

The Bohm-de Broglie (BdB) interpretation of homogeneous minisuperspace models
can be summarized
as follows\cite{bola}: the Wheeler-DeWitt equation is

\begin{equation}\label{25}
{\cal H}[\hat{p}^{\alpha}(t), \hat{q}_{\alpha}(t)]\Psi(q)=0.
\end{equation}
Substituting the wave function in polar form, $\Psi=Ae^{iS}$, we have a
complex equation and its
real part produce, after dividing it by $A$:

\begin{equation}\label{26}
\frac{1}{2}f_{\alpha \beta}(q_{\mu})\frac{\partial S}
{\partial q_{\alpha}}\frac{\partial S}{\partial q_{\beta}}+U(q_{\mu})+Q(q_{\mu})=0 ,
\end{equation}
where $ Q(q_{\mu})$ is the quantum potential, given by

\begin{equation}\label{27}
Q(q_{\mu})=-\frac{1}{2 A} f_{\alpha \beta}\frac{\partial^2 A}
{\partial q_{\alpha} \partial q_{\beta}} .
\end{equation}

In the BdB approach, the trajectories $q_{\alpha}(t)$
are supposed
to be real, independent of any observations.
Equation (\ref{26}) is the Hamilton-Jacobi equation for them, but with an extra
term given by the quantum potential (\ref{27}). Then we define

\begin{equation}\label{28}
p^{\alpha}=\frac{\partial S}{\partial q_{\alpha}} ,
\end{equation}
where the momenta are related to the velocities in the usual way:

\begin{equation}\label{29}
p^{\alpha}=f^{\alpha \beta}\frac{1}{N}\frac{\partial q_{\beta}}{\partial t}.
\end{equation}
In order to obtain the quantum trajectories we have to solve the Bohm
guidance relations,
$p=\frac{\partial S}{\partial q}$, which are, in this case, given by:

\begin{equation}\label{30}
\frac{\partial S(q_{\alpha})}{\partial q_{\alpha}}=
f^{\alpha \beta}\frac{1}{N}\frac{\partial q_{\beta}}{\partial t}.
\end{equation}
Equations (\ref{30}) are invariant under time reparametrization. Hence, even at
the quantum level,
different choices of $N(t)$ yield the same spacetime geometry for a given
nonclassical
solution $q_{\alpha}(t)$. There is no problem of time in BdB interpretation
of minisuperspace quantum cosmology\footnote{ This is not true for the full
superspace, see \cite{must}, although the theory remain consistent,
see \cite{cons}, \cite{tese}.}.

In the case of hamiltonian (\ref{9}), we obtain for the guidance relations
(\ref{30})

\begin{equation}\label{31}
\frac{\partial S}{\partial \alpha}=-\frac{e^{3\alpha}\dot{\alpha}}{N} ,
\end{equation}

\begin{equation}\label{32}
\frac{\partial S}{\partial \phi}=\frac{e^{3\alpha}\dot{\phi}}{N} .
\end{equation}
The modified Hamilton-Jacobi equation (\ref{26}) reduces to

\begin{equation}\label{hj}
\frac{1}{2}\biggr[\bigg(\frac{\partial S}{\partial \phi}\biggl)^2 -
\biggr(\frac{\partial S}{\partial \alpha}\biggl)^2   \biggl]   -
\frac{\epsilon}{2}
e^{4\alpha}-\frac{1}{2A}\biggr(\frac{\partial^2 A}{\partial \phi^2}-\frac{\partial^2 A}{\partial \alpha^2}\biggl)=0 ,
\end{equation}
where the last term in the LHS represents the quantum potential (\ref{27}):

\begin{equation}\label{33}
Q(\alpha,\phi)\equiv
\frac{1}{2 A}\biggr[ \frac{\partial^2 A}{\partial \alpha^2}-
\frac{\partial^2 A}{\partial \phi^2}\biggl] .
\end{equation}

We will now apply the BdB interpretation to our
minisuperspace model.
We will restrict ourselves to the case of flat spatial sections,
hypersurfaces with $\epsilon=0$.
In Ref. \cite{fab2}, the following Gaussian
superpositions of the solutions (\ref{20}) were studied:

\begin{equation}\label{34}
\Psi_{1}(\alpha,\phi)=\int F(k)B_k(\phi)[A_k(\alpha)+A_{-k}(\alpha)]dk ,
\end{equation}
and

\begin{equation}\label{35}
\Psi_{2}(\alpha,\phi)=\int F(k)A_k(\alpha)[B_k(\phi)+B_{-k}(\phi)]dk ,
\end{equation}
both with $a_2=b_2=0$ in Eqs (\ref{21},\ref{22}),
and where the arbitrary function $F(k)$ is the Gaussian

\begin{equation}\label{36}
F(k)=\exp{\biggr[-\frac{(k-d)^2}{\sigma^2}}\biggl].
\end{equation}
While in paper \cite{fab2} the solution $\Psi_1$  was studied in detail,
here we will concentrate our analysis on the solution $\Psi_2$.
Integrating (\ref{36}) in $k$ we obtain, for $\Psi_2$,

\begin{eqnarray}\label{37}
&\Psi_2=a_1 b_1 \mid \sigma \mid \sqrt{\pi}\biggr\{\exp{\biggr
[ -\frac{(\alpha+\phi)^2\sigma^2}{4}\biggl]} &
\exp(id(\alpha+\phi)) \nonumber \\
&+\exp{\biggr[ -\frac{(\alpha-\phi)^2\sigma^2}{4}\biggl]}
\exp(id(\alpha-\phi))\biggl\}&
\end{eqnarray}
To obtain the quantum trajectories, we have to calculate the phase $S$ of
the above wave function
and substitute it into the guidance equations (\ref{31}) and (\ref{32}).
We will work in the
gauge $N=1$. Computing the phase of $\Psi_2$, we obtain
$S=d\alpha +\arctan(\frac{\sigma^2 \phi \alpha}{2}) \tan(-d\phi)$ which,
after substitution in Eqs (\ref{31},\ref{32}), yields a planar system given by:

\begin{equation}\label{38}
\dot{\alpha}=\frac{\phi \sigma^2 \sin(2d\phi)-2d \cos(2d\phi)-2d \cosh(\sigma^2\alpha\phi)}
{e^{3\alpha}2[\cos(2d\phi)+\cosh(\sigma^2\alpha\phi)]}
\end{equation}

\begin{equation}\label{39}
\dot{\phi}=-\frac{\alpha\sigma^2 \sin(2d\phi)+2d \sinh(\sigma^2\alpha\phi)}
{e^{3\alpha}2[\cos(2d\phi)+\cosh(\sigma^2\alpha\phi)]}
\end{equation}
Equations (\ref{38},\ref{39}) give the direction of the geometrical tangents
to the trajectories which solves this planar system.
By plotting the tangent direction field, it is possible to obtain the trajectories.
The vertical line $\phi=0$ divides the configuration space in two symmetric
regions.
The line $\alpha=0$ contains all singular points of this system, which are nodes
and centers. The
nodes appear when the denominator of the above equations, which is proportional to
the norm
of the wave equation, is zero. No trajectory can pass through these points. They
happen
when $\alpha=0$ and $\phi=(2n+1)\frac{\pi}{2d}$, $n$ an integer, with periodicity
$\pi/|d|$.
The center points appear when the numerators are zero.  They are given by
$\alpha=0$ and
$\phi = \frac{2d}{\sigma^2}\cot(d \phi)$.
As $|\phi| \rightarrow \infty$ these points tend to $n \pi /d$ (zeros of
$\tan(d \phi)$).
As one can see from the above system, the classical solutions
$(a(t) \propto t^{1/3})$ are recovered when $|\phi| \rightarrow \infty$
or $|\alpha| \rightarrow \infty$, the other being different from zero.

\begin{figure}
\includegraphics{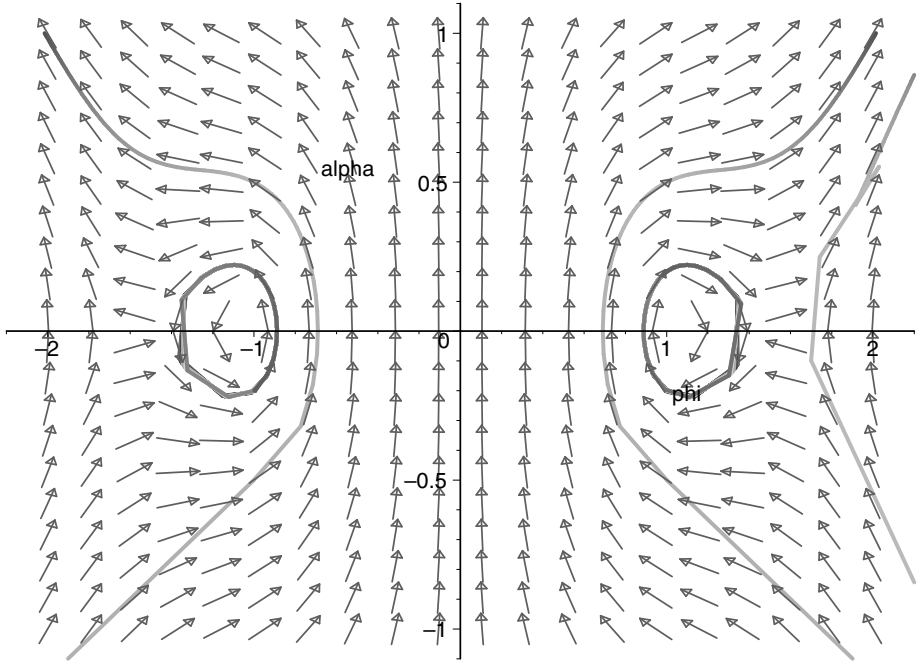}
\caption {Field plot of the system of planar equations 
(\ref{38},\ref{39}) coming
from the wave function (\ref{37}).For numerical simplicity we choose 
the values $-d=\sigma=1$. We see the two possibilities: trajectories
corresponding to
oscillating universes without singularities and trajectories corresponding
to non-oscillating universes that come from a singularity, experience
quantum effects in the middle of their
expansions, and recover their classical behaviours for large values of $\alpha$.}
\label{configu}
\end{figure}

We present a field plot of this planar system in Fig. \ref{configu}, for the case $d=-1$,
$\sigma=1$.
Depending on the initial conditions, we can see two different possibilities.
Near the center points there are oscillating universes without
singularities and with amplitudes of oscillation of order 1.
The other possibility is given by non-oscillating universes.
A non oscillating universe arises classically from a singularity, experiences
quantum effects
in the middle of its expansion, and recover  its classical behaviour for large
values of $\alpha$.


\section{The  accelerated expansion.}
\subsection{Qualitative approach}
In this subsection we show how our approach for the explanation of the 
accelerated expansion works in a qualitative manner, i.e.,
with  the parameters of the wave packet ($d$ and $\sigma$) adapted 
for readable numerical treatment, without any fitting with usual cosmological
orders of magnitude. The aim is just to show that in some period in the
history of such quantum models where the expansion is positively accelerated.
We take here $d=-1$,
$\sigma=1$ for the
numerical computations. 
The quantum effects appearing in the middle of the non periodic bohmian
trajectories described above can deviate them from their classical decelerated
expansion to an accelerated one. We will show that this is indeed the case
of this model. From $\alpha=\ln{a(t)}$ we have

\begin{equation}\label{40}
\frac{\ddot{a}}{a}=\ddot{\alpha}+(\dot{\alpha})^2.
\end{equation}
From Eq. (\ref{38}), $\dot{\alpha}$ can be viewed as a function of the canonical
variables
$\dot{\alpha}=f(\alpha,\phi)$. Then we have

\begin{equation}\label{40a}
\ddot{\alpha}=\frac{\partial f}{\partial \alpha}\dot{\alpha}+\frac{\partial f}{\partial \phi}\dot{\phi}
\end{equation}
Computing the derivatives with respect to $\alpha$ and $\phi$, and substituting
$\dot{\alpha}$
and $\dot{\phi}$ from equations (\ref{38}) and (\ref{39}), respectively,
we obtain





\begin{widetext}
\begin{eqnarray}\label{41}
\frac{\ddot{a}}{a}=-\frac{1}{4}\biggr[ \left (2{\phi}^{2}{\sigma}^{4}+\alpha{\sigma}^{4}
\right )\left (\left (\sin(2d\phi)\right )^{2}\cos(2d\phi)+\left (
\sin(2d\phi)\right )^{2}\cosh({\sigma}^{2}\alpha\phi)\right )+ \nonumber \\
\left (2{\sigma}^{2}d-2{\sigma}^{4}d{\phi}^{2}\right )\left (\sin(
2d\phi)\cos(2d\phi)\sinh({\sigma}^{2}\alpha\phi)+\sin(2d\phi)
\sinh({\sigma}^{2}\alpha\phi)\cosh({\sigma}^{2}\alpha\phi)\right )+\nonumber \\
\left ({\sigma}^{6}{\phi}^{3}+4{\sigma}^{2}{d}^{2}\phi-{\sigma}^{6}
\phi{\alpha}^{2}\right )\left (\sin(2d\phi)\right )^{2}\sinh({
\sigma}^{2}\alpha\phi)+
\left (2{\sigma}^{4}d\phi\alpha-8{
\sigma}^{2}d\phi\right )\sin(2d\phi)\left (\cos(2d\phi)\right )^{2
}+\nonumber \\
\left (2{\sigma}^{4}d\phi\alpha-16{\sigma}^{2}d\phi\right )
\sin(2d\phi)\cos(2d\phi)\cosh({\sigma}^{2}\alpha\phi)+
24{d}^{2
}\left (\cos(2d\phi)\right )^{2}\cosh({\sigma}^{2}\alpha\phi)+\nonumber \\
24{d}^{2}\cos(2d\phi)\left (\cosh({\sigma}^{2}\alpha\phi)\right )^{2
}-
8\phi d{\sigma}^{2}\sin(2d \phi)\left (\cosh({\sigma}^{2}\alpha
\phi)\right )^{2}+4{\sigma}^{2}\phi{d}^{2}\left (\cos(2d\phi)
\right )^{2}\sinh({\sigma}^{2}\alpha\phi)+\nonumber \\
4{\sigma}^{2}\phi{d}^{
2}\cos(2d\phi)\cosh({\sigma}^{2}\alpha\phi)\sinh({\sigma}^{2}
\alpha\phi)+2{\sigma}^{4}\phi\left (\sin(2d\phi)\right )^{3}d
\alpha-
2{\sigma}^{4}\phi\sin(2d\phi)\left (\sinh({\sigma}^{2}
\alpha\phi)\right )^{2}\alpha d+\nonumber \\
8{d}^{2}\left (\cos(2d\phi)
\right )^{3}+
8{d}^{2}\left (\cosh({\sigma}^{2}\alpha\phi)\right )^
{3}\biggl]
\biggr[\left (e^{3\alpha}\right )^{2}\left (\cos(2d\phi)+\cosh({
\sigma}^{2}\alpha\phi)\right )^{3}\biggl]^{-1} .
\end{eqnarray}
\end{widetext}
The  equation above gives the  acceleration
$\frac{\ddot{a}}{a}$ as a function of $\alpha$ and $\phi$. If one integrates the system
(\ref{38},\ref{39}) to obtain $\phi= \phi(a)$, the quantum version of the classical 
equation (i.e. Raychaudhuri equation for the Friedmann model)

\begin{equation}
\label{ray}
\frac{\ddot{a}}{a}=-\frac{4 \pi G}{3} (\rho+3 p) \propto -(\dot{\phi})^2
\propto -\frac{1}{a^6}
\end{equation}
can be obtained.
Taking the limit where the absolute value of $\alpha$ is very large
in Eq. (\ref{41}),
one recovers the classical behaviour given in Eq. (\ref{ray}),
$\frac{\ddot{a}}{a}\propto -1/a^6$. Note that the quantum analog of the
Friedmann's equation can also be easily obtained
from Eqs.(\ref{31},\ref{32},\ref{hj}), yielding (recovering the unities)
$H^2={\dot{\phi}}^2+\beta ^4c^2Q(a,\phi)/a_{\rm phys}^6$.

\begin{figure}\includegraphics{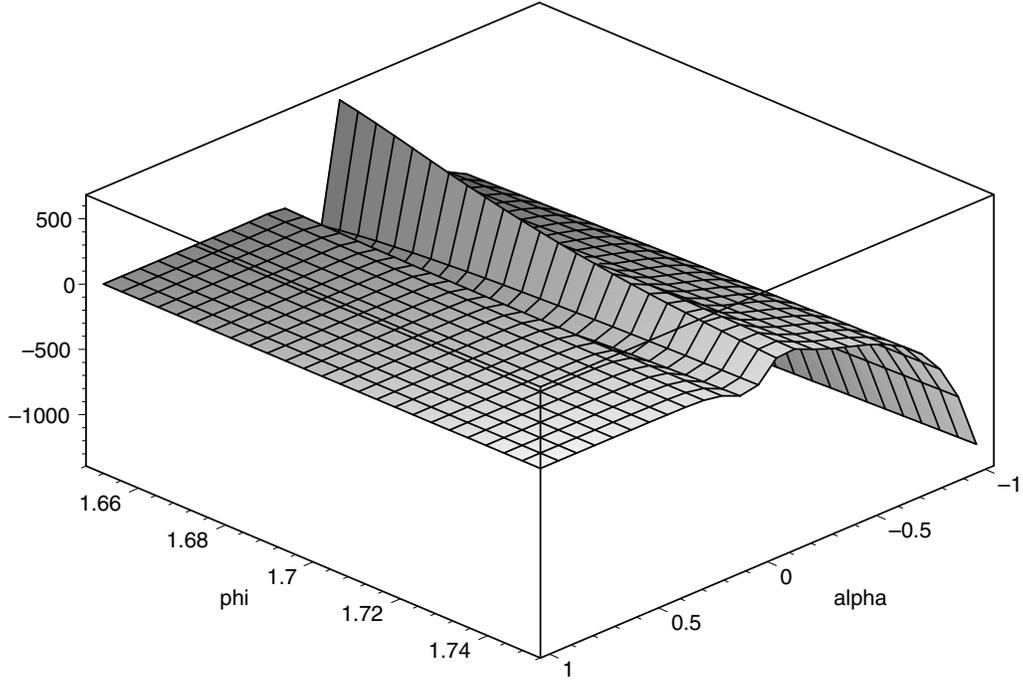}
\caption{Acceleration $\ddot{a}/a$ as a function of $\phi$ and $\alpha$.
For numerical simplicity we choose the values $-d=\sigma=1$.}
\label{ace}
\end{figure}

\begin{figure}\includegraphics{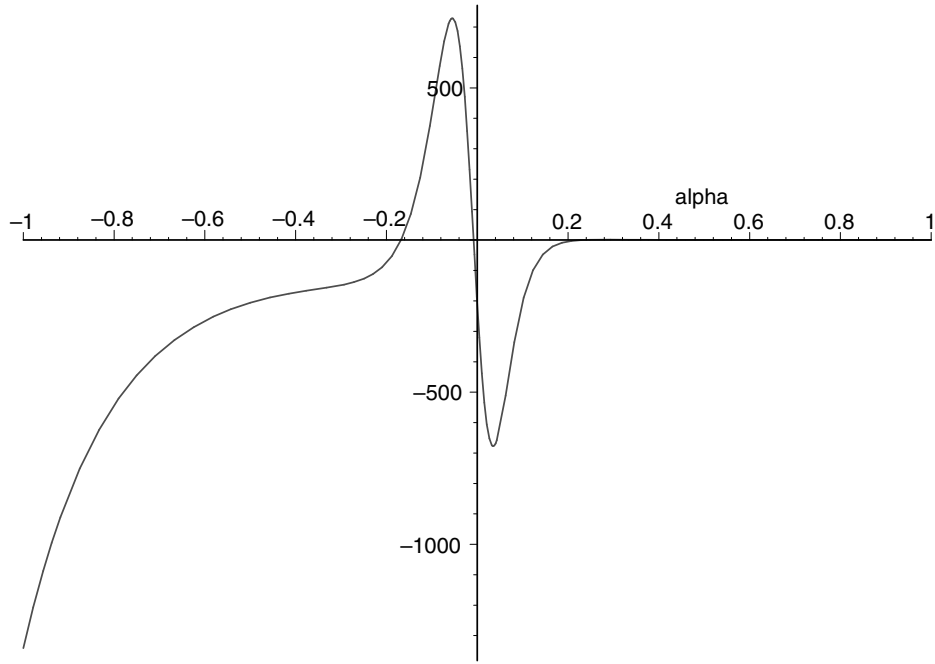}
\caption{Acceleration $\ddot{a}/a$ as a function of $\alpha$
for $\phi=1.66$. For numerical simplicity we choose the values $-d=\sigma=1$.}
\label{detalhe1}
\end{figure}

We can represent $\frac{\ddot{a}}{a}$ in a tridimensional
plot as a function of $\alpha$ and $\phi$.
In this plot we can see the regions on
the plane $\alpha-\phi$ in which the acceleration is possitive, negative or zero.
We show this plot, for the parameters $d=-1$ and $\sigma=1$, in Fig. \ref{ace}.
One can see the classical behaviour $\ddot{a}/a
 \propto -1/a^6
$ for $ a \rightarrow 0$ ($\alpha \rightarrow -\infty$) and
$a \rightarrow \infty$ ($\alpha \rightarrow \infty$), but near the region
$a=1$ ($\alpha=0$), a clear departure from classical
behaviour is observed, and positive values of $\ddot{a}/a$ are obtained.
Fig. \ref{detalhe1} shows the acceleration $\ddot{a}/a$ as a function of $\alpha$ for
$\phi=1.66$.

\begin{figure}
\includegraphics{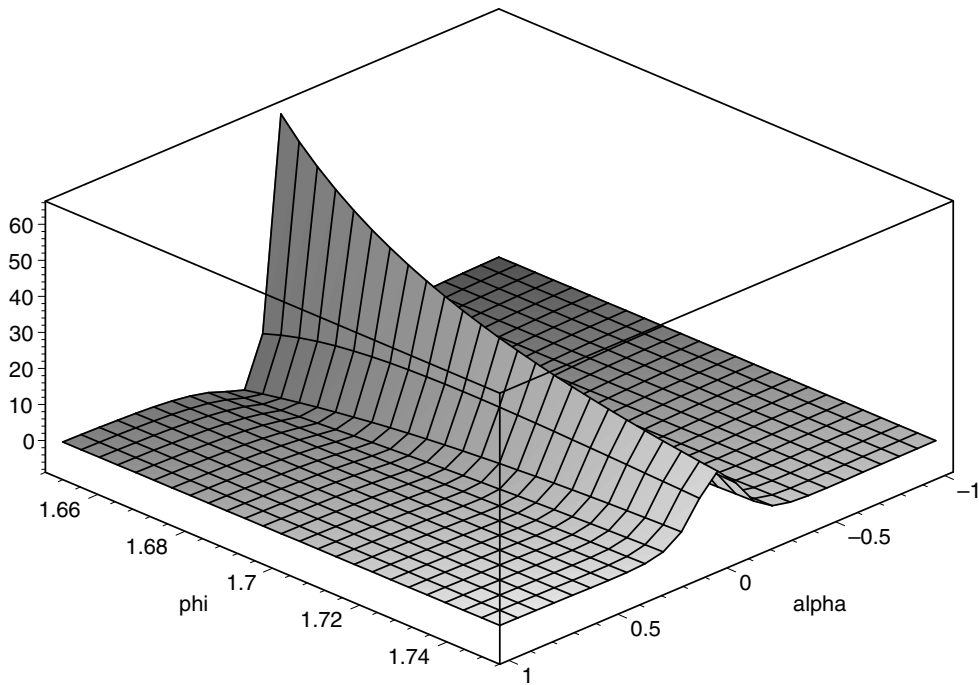}
\caption{The quantum potential as a function of  $\phi$ and $\alpha$. We
see that it is different from zero in the region of positive acceleration. For
numerical simplicity we choose the values $-d=\sigma=1$.}
\label{potqua}
\end{figure}

\begin{figure}
\includegraphics{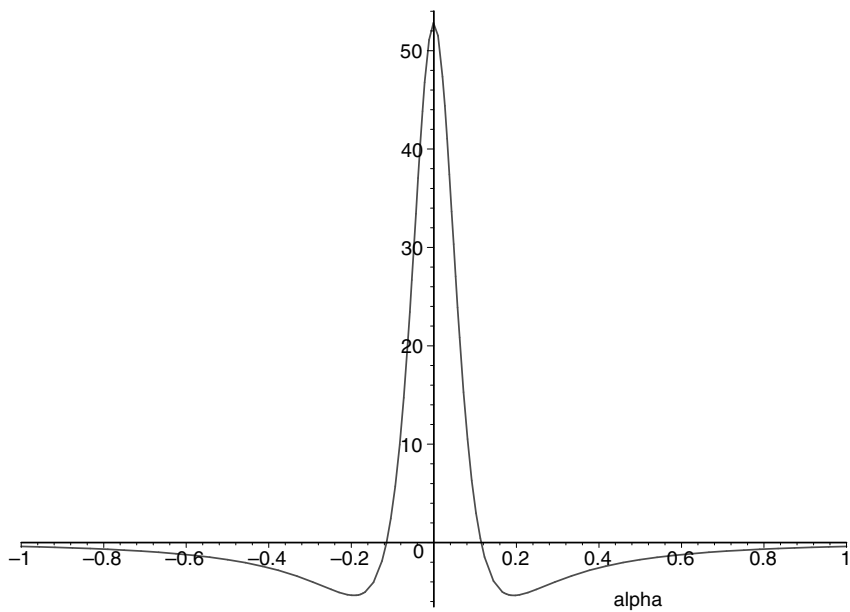}
\caption{The quantum potential as a function of $\alpha$ for $\phi=1.66$. We
see that it is different from zero in the region of positive acceleration. For
numerical simplicity we choose the values $-d=\sigma=1$.}
\label{detalhe2}
\end{figure}

As we pointed above, the quantum
potential is the cause of this positive acceleration.
Fig. \ref{potqua} shows the quantum potential in the $\alpha-\phi$ plane
where we can see
that it is different from zero in the regions of positive acceleration.
A trajectory passing through this region on the plane $\alpha-\phi$ will
correspond to a universe experiencing an accelerated expansion.
Fig. \ref{detalhe2} shows the quantum potential as a function of $\alpha$ for
$\phi=1.66$. Note also the increasing in the acceleration and in the quantum
potential as we approach the node point $\alpha =0$, $\phi=\pi/2$.


\subsection{Comparison between the quantum acceleration of the Universe's expansion and the presence
of a cosmological constant}

We will now compare quantitatively the accelerated expansion of the Universe caused by such a quantum  
cosmological effect  with  the one generated by the presence of a cosmological constant in the classical 
model . First of all, we must recover the units in Eqs.(\ref{38},\ref{39}).
This is done by multiplying the
RHS of these equations  by $c/\beta$, where
$\beta=\sqrt{\frac{4\pi}{3V}}l_P$ and $V=V_{{\rm phys}}(t)/a^3_{{\rm phys}}(t)$,
where $ a_{{\rm phys}}=a\beta$ is the physical  scale factor and
$ V_{{\rm phys}}$ is the physical volume of the spacelike section. Then we obtain

\begin{equation}\label{qcalfa}
H=\dot{\alpha}=f(\alpha,\phi)\frac{V_P}{V_{{\rm phys}}(t) t_P} ,
\end{equation}

\begin{equation}\label{qcfi}
\dot{\phi}=g(\alpha,\phi) \frac{V_P}{V_{{\rm phys}}(t) t_P} ,
\end{equation}
where

\begin{equation}\label{38beta}
f(\alpha,\phi)=\frac{\phi \sigma^2 \sin(2d\phi) }
{2 [\cos(2d\phi)+\cosh(\sigma^2\alpha\phi)]}-d ,
\end{equation}

\begin{equation}\label{39bis}
g(\alpha,\phi)=-\frac{\alpha\sigma^2 \sin(2d\phi)+2d \sinh(\sigma^2\alpha\phi)}
{2[\cos(2d\phi)+\cosh(\sigma^2\alpha\phi)]} ,
\end{equation}
$V_P\equiv\frac{4\pi}{3} l^3_P $ is the Planck volume, and $t_P$ is the Planck time.
We will compare this quantum cosmological model with the original classical free scalar 
field model, classically equivalent to stiff matter, with flat spatial section, suplemented 
with a cosmological constant as an alternative source for accelerated expansion. This classical
model satisfies the Friedmann´s equation

\begin{equation}
H^2=\frac{8\pi G}{3}\frac{c_{\phi}}{a^6}+\Lambda=H_0^2
[(1-\Omega_{\Lambda})(1+z)^6+\Omega_{\Lambda}] ,
\end{equation}
where $c_{\phi}$ is a constant such that the energy density of 
the field is $\rho_{\phi}=c_{\phi}/a^6$,
$\Omega_{\Lambda}=\Lambda/(8\pi G \rho_{\rm crit})$, $\rho_{\rm crit}$
is the critical density,
 and $H_0$ is the Hubble's paremeter today.
 The deceleration parameter today, $q_0=-\ddot{a}/(a{H_0^2})$, is given by

  \begin{equation}\label{ccos}
q_0=2-3\Omega_{\Lambda}.
 \end{equation}
 To obtain the luminosity distance as a function of $z$ one can calculate numerically the integral

 \begin{equation}\label{dlum}
d_L=(1+z)\int_{0}^{z}\frac{{\rm d}y}{H(y)}=\frac{(1+z)}{H_0}\int_{0}^{z}
\frac{{\rm d}y}{\sqrt{(1-\Omega_{\Lambda})(1+y)^6+\Omega_{\Lambda}}} .
\end{equation}
As a power series, it can be written as

\begin{equation}\label{dlum2}
H_{0}d_L=z+z^2\biggl(-\frac{1}{2}+\frac{3}{2}\Omega_{\Lambda}\biggr)+
z^3(1-\Omega_{\Lambda})\biggl(\frac{1}{2}-\frac{9}{2}\Omega_{\Lambda}\biggr)+... ,
\end{equation}

In the quantum cosmological problem, we have to deal with Eqs.
(\ref{qcalfa},\ref{qcfi}). There are four arbitrary parameters: $\sigma, d$
and two
coordinates in the $(\alpha,\phi)$ plane, $\phi_0$ designating the particular
trajectory in Fig.(\ref{configu}) which represent
our Universe, and $\alpha_0$ designating
the present moment in a particular trajectory. The present value of the
Hubble's paremeter
in a particular trajectory coming from Eq.(\ref{qcalfa}) is given by

\begin{equation}\label{part}
H_0=f(\alpha_0,\phi_0)
\frac{V_P}{V_{{\rm phys}}^{0} t_P}
\end{equation}
where

\begin{equation}\label{fhoje}
f(\alpha_0,\phi_0) \equiv
\frac{\phi_0 \sigma^2 \sin(2d\phi_0)}
{2 [\cos(2d\phi_0)+\cosh(\sigma^2\alpha_0\phi_0)]}-d
\end{equation}
In order to obtain a model similar to our present Universe
whith $H_0 \simeq 10^{-18}{\rm s}^{-1}$ and
$V_{{\rm phys}}^{0}\geq 10^{82}{\rm cm}^3$,
one must have

\begin{equation}\label{fhoje2}
f(\alpha_0,\phi_0) \equiv
\frac{\phi_0 \sigma^2 \sin(2d\phi_0)}
{2 (\cos(2d\phi_0)+\cosh(\sigma^2\alpha_0\phi_0))}-d\simeq 10^{120}
\end{equation}
This huge number can be obtained by choosing a very large value for $\sigma^2$
(the gaussian in the wave function would
be almost flat indicating no preference in the choice of $k$, or physically,
no prefered choice in the strength of the initial explosion),
a very large value for $|d|$ (a gaussian
centered in a very negative value of $k$, or a choice for a very strong initial
explosion), or trajectories passing very close to the node point,
where the denominator of the above expression
 approaches zero.

 In the case of a large value of $\sigma^2$, choosing
 $\alpha_0 \simeq 1/\sigma^4 \simeq 0$,
 one can check from Eq.(\ref{41}) that

\begin{equation}
q_0=-\frac{\ddot{a}}{a}|_0\frac{1}{H_0^2}\simeq
\frac{2\cos(2d\phi_0)+1}{\cos(2d\phi_0)+1}
\end{equation}.
For $2d\phi_0=2n\pi+x$
with $x \in (2.145,2.15)$ one has $-0.21<q_0<-0.17$. Note that,
for the classical model
with $\Omega_{\Lambda}=0.73$ \cite{new},  $q_0=-0.19$ (see Eq. (\ref{ccos})).

The supernovae measurements relate the luminosity distance $d_L$ with $z$.
Hence, it would be instructive to compare the quantum cosmological
luminosity distance $d_L^q(z)$

\begin{equation}\label{dquan}
d_L^q=(1+z)\int_0^z\frac{{\rm d}y e^{3\alpha(y)}}{f(\alpha(y),\phi(y))}
\frac{\beta}{c} ,
\end{equation}
with the one given in Eqs.(\ref{dlum},\ref{dlum2}).
We will expande $G(y)\equiv e^{3\alpha(y)}/f(\alpha(y),\phi(y))$ in powers of
$y$ around $y=0$ (today)

\begin{equation}\label{gex}
G(y)=G(0)+y\frac{{\rm d}G}{{\rm d}y}|_{y=0}+\frac{y^2}{2!}
\frac{{\rm d}^2G}{{\rm d}y^2}|_{y=0}+
\frac{y^3}{3!}\frac{{\rm d}^3G}{{\rm d}y^3}|_{y=0}+...
\end{equation}
up to third order.
The operator $ {\rm d}/{\rm d}y $ can be written as

\begin{equation}\label{oper}
\frac{{\rm d}}{{\rm d}y}=-\frac{1}{(1+y)}\frac{{\rm d}}{{\rm d}\alpha}=
-\frac{1}{(1+y)}\biggl(\frac{\partial}{\partial \alpha}+
\frac{{\rm d}\phi}{{\rm d}\alpha}\frac{\partial}{\partial \phi}\biggr) ,
\end{equation}
where ${\rm d}\phi/{\rm d}\alpha$ can be easily obtained from
Eqs.(\ref{38},\ref{39}) and $\alpha = \ln(a) = -\ln(1+y) + {\rm const.}$.
Recalling that

\begin{equation}
\frac{H_0\beta a_0^3}{c}=\frac{H_0 a_{\rm phys}^3}{c \beta^2}=\frac{H_0 a_{\rm phys}^3
l_P 3 V}{4\pi}=\frac{H_0 t_P V_{\rm phys}}{V_P} \simeq 10^{120} ,
\end{equation}
we obtain

\begin{eqnarray}\label{dquan2}
H_0 d_L^q &=& \frac{10^{120}}{f_0}\biggl[z+z^2\biggl(1+\frac{f_0}{2a_0^3}
\frac{{\rm d}G}
{{\rm d}z}|_0 \biggr)+
z^3\biggl( \frac{f_0}{2a_0^3}\frac{{\rm d}G}{{\rm d}z}|_0+\frac{f_0}{6 a_0^3}
\frac{{\rm d}^2G}{{\rm d}z^2}|_0\biggr)+ \nonumber \\
& & + z^4\biggl(\frac{f_0}{6a_0^3}\frac{{\rm d}^2G}{{\rm d}z^2}|_0 +
\frac{f_0}{24a_0^3}\frac{{\rm d}^3G}{{\rm d}z^3}|_0\biggr)+...\biggr] .
\end{eqnarray}

Comparing Eq.(\ref{dquan2}) with Eq.(\ref{dlum2}), one can see that $f_0$ must be of
order $10^{120}$, as we have already concluded in Eq. (\ref{fhoje}).
Here we will choose $|d|$ as the big number to make $f_0 \simeq 10^{120}$,
and $|\alpha_0|=\gamma/|d|$, $\gamma$ being an arbitrary number of
order $1$.
This is to assure that $\frac{{\rm d}^n G}{{\rm d}z^n}|_0$ is of order $1$ or less
for any $n \geq 1$,
which is not the case if we choose $\sigma \gg 1$\footnote{In this case,
the series expansion of Eq.(\ref{dquan}) is not meaningful at lower orders, and the
integral must be performed by other methods.}.
Using Eq.(\ref{oper}), we can calculate the coefficients in Eq.(\ref{dquan2}),
obtaining

\begin{eqnarray}\label{dquan3}
H_0 d_L^q &\simeq& z-\frac{1}{2}z^2+z^3\biggl\{\frac{1}{2}+\frac{ \phi_0^2 \sigma^4}
{ 6[1+ \cos(2d \phi_0)]^2 }\biggr\} + \nonumber \\
& &+z^4\biggl\{-\frac{1}{2}- \frac{ \phi_0^2 \sigma^4}{3[1+\cos(2d\phi_0)]^2}-
 \frac{\phi_0^3 \sigma^6 \gamma \sin(2d\phi_0)}{2[1+\cos(2d \phi_0)]^4}\biggr\} +...
\end{eqnarray}
In order to obtain Eq.(\ref{dquan3}), we have used that $|d|>>1$ and
$|\alpha_0|=\gamma/|d|$ is very small\footnote{Note that the first correction
to the classical model without cosmological constant (see Eq.(\ref{dlum2})
with $\Omega_{\Lambda}=0$) come in the cubic term. The correction to the
deceleration parameter at this particular moment is negligible with this
choice of parameters. However, the corrections in the cubic and forth terms
can be adjusted in order to make the curve obtained from Eq.(\ref{dquan3})
close to the corresponding curve obtained from Eq.(\ref{dlum}),
as we will see.}.

There are many values of $\phi_0, \sigma_0$ and $\gamma$ which makes the
graphic of $H_0 d_L^q$  similar to the one obtained from Eq. (\ref{dlum})
for $\Omega_{\Lambda}=0.73$. For instance, for $\gamma=1$ we set
$2d\phi=0.64\pm 2n\pi$ and the coefficients of $z^3$ and $z^4$ become,
respectively, $2.6$ and $-2.25$.

\begin{figure}
\includegraphics{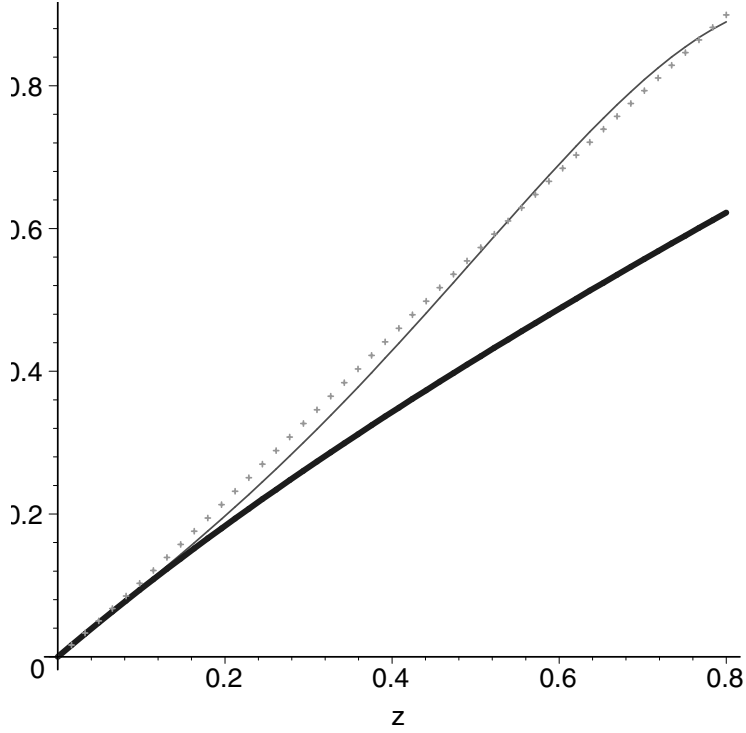}
\caption{The luminosity distance as a function of  redshift. The thin line
curve corresponds to the quantum model, the dotted curve is for the
classical model with a cosmological constant, and the thick line curve
is for the classical model without cosmological constant.}
\label{compare2}
\end{figure}

In Fig.\ref{compare2}, we show a plot of $H_0 d_L(z)$ given by Eq.(\ref{dlum}),
$H_0 d_L(z)$ given by Eq.(\ref{dlum}) with $\Omega_{Lambda}=0$, and $H_0 d_L^q(z)$
given by
Eq.(\ref{dquan3}) with the coefficients of $z^3$ and $z^4$ being equal to $2.6$
and $-2.25$, respectively. Note that for small values of  $z$ they are close but,
for intermediary values of $z$, the quantum $d_L^q(z)$ remain close to the
cosmological constant $d_L(z)$ while both separates of the pure stiff matter
$d_L(z)$. Of course, for bigger values of $z$, the quantum $d_L^q(z)$ may separate
strongly from the cosmological constant $d_L(z)$. Hence, quantum  comological
effects may mimic a cosmological constant in some region but not everywhere.
The two models are distinguishable.

\section{Conclusion}
We have studied  gaussian superpositions of positive and negative frequency
mode solutions of the Wheeler-DeWitt equation corresponding to
a scalar-tensor model in minisuperspace in the case of flat spatial section.
According to the Bohm-de Broglie
approach to quantum cosmology, the quantum
trajectories representing dynamical universes evolving
in time were studied.
We have shown that it is possible to have universes which arise classically
from a singularity, undergo a positive acceleration
in the middle of its expansion, and recover  its classical behaviour for large
values of the scale factor. We have shown that this positive acceleration, which
can be made compatible with observations for many choices of initial
conditions, is due to a quantum cosmological effect driven
by the quantum potential, according to the BdB interpretation of quantum cosmology.
In this way, it may be possible to explain the positive acceleration
suggested by the recent measurements of high redshift supernovae
without postulating a new contribution to the energy density of the Universe as
the dark energy. Note that this acceleration is caused by quantum effects not
only present in
the scalar field, as described in Ref.\cite{parker}, but also in the geometry
itself. We consider the Universe as a quantum system
no matter its size. It is possible to have small classical
universes and large quantum ones:  it depends on the  state functional and on
initial conditions \cite{fab,hartle}.

The quantum cosmological explanation for the acceleration of the Universe
presented in this paper needs to
be studied further, not only because it would be an alternative explanation for a 
misterious behaviour of the present Universe without appealing to any new form of
energy, but also
because it is a possibility of an observable physical effect of quantum cosmology.
Furthermore, quantum cosmological explanations may be suported by
symmetry principles which are absent in the classical domain. As an example,
we have seen that
the huge cosmological numbers may be explained by some "democracy" principle
stating that any value of the velocity of expansion (the constant
$k$) is equally good ($\sigma$ is very big, the gaussian is almost flat).
Of course, more
elaborated models taking into account relevant matter sources like dust and
radiation
must be studied. This will
be the subject of our future investigations.

\section*{ACKNOWLEDGEMENTS}
We would like to thank {\it Conselho Nacional de Desenvolvimento Cient\'{\i}fico e
Tecnol\'ogico}
(CNPq) of Brazil
for financial support.
ESS was succesively supported  by a postdoctoral FACITEC ({\it
Fundo de Apoio \'a Ci\^encia e Tecnologia
do Munic\'{\i}pio de Vit\'oria -- ES}) fellowship, a
visiting CBPF  fellowship, and a
    posdoctoral CLAF-CNPq  fellowship. We want to thank Sonia Ferreira and
    Zelia Quadros from Lafex-CBPF for technical support.
We would also like to thank `Pequeno Seminario' of CBPF's cosmology
group
for useful discussions.
    \vspace{1.0cm}



\begin{thebibliography}{99}

\bibitem{fab2} R. Colistete Jr., J. C. Fabris and N. Pinto-Neto,
Phys. Rev. {\bf D62}, 83507 (2000).

\bibitem{SN1} S. Perlmutter, {\it et al.}, Nature (London) {\bf 391}, 51 (1998).

\bibitem{SN} A. Riess {\it et al.}, Astron. J. {\bf 116}, 1009, (1998).

\bibitem{triangle} Neta A. Bahcall, J. P. Ostriker, S. Perlmutter and
P. J. Steinhardt, Science 284, 1481-1488 (1999).

\bibitem{garnavich} P.M. Garnavich {\it et al.},
Astrophys.J. {\bf 509},74 (1998).

\bibitem{wang} L. Wang,  R.R. Caldwell, J.P. Ostriker and Paul J. Steinhardt
Astrophys. J. {\bf 530}, 17-35 (2000).

\bibitem{quint} R. R. Caldwell, R. Dave and P. J. Steinhardt,
Phys. Rev. Lett.{\bf 80}, 1582 (1998);

\bibitem{jer} Philippe Brax and Jerome Martin, Phys.Rev. {\bf D61}
103502 (2000).

\bibitem{fab} R. Colistete Jr., J. C. Fabris and N. Pinto-Neto,
Phys. Rev. {\bf D57}, 4707 (1998).

\bibitem{bohm1} David Bohm,  Phys. Rev. {\bf 85}, 166 (1952).

\bibitem{bohm2} David Bohm,  Phys. Rev. {\bf 85}, 180 (1952).

\bibitem{hol} P. R. Holland,
{\it The Quantum Theory of Motion: An Account of the de Broglie-Bohm
Causal Interpretation of Quantum Mechanics}
(Cambridge University Press, Cambridge, 1993).

\bibitem{eve} B. S. DeWitt and N. Graham (Eds.)
{\it The Many-Worlds Interpretation of Quantum Mechanics}
(Princeton University Press, Princeton, 1973).

\bibitem{hartle} Jonathan J. Halliwell and James B. Hartle, Phys. Rev.
{\bf D 41}, 1815 (1990).

\bibitem{parker} Leonard Parker and Alpan Raval, Phys. Rev. Lett.
{\bf 86}, 749 (2001).

\bibitem{dl} V. A. De Lorenci, J. Martin, N. Pinto-Neto and I. Dami\~ao Soares,
Phys.Rev. {\bf D 56}, 3329 (1997).

\bibitem{dirac} P. A. M. Dirac, {\it Lectures on Quantum Mechanics},
Yeshiva University (1964).

\bibitem{bola} J. Ac\'acio de Barros and N. Pinto-Neto,
Int. J. of Mod. Phys. {\bf D 7}, 201 (1998).

\bibitem{must} N. Pinto-Neto and E. Sergio Santini, Phys.Rev. {D \bf 59}
123517 (1999).

\bibitem{cons} N. Pinto-Neto and E. Sergio Santini, Gen. Rel. and Grav. {\bf 34},
505 (2002).

\bibitem{tese} E. Sergio Santini, PhD Thesis, CBPF-Rio de Janeiro, (May 2000),
(gr-qc/0005092).

\bibitem{new} C. L. Bennett {\it et al.}, astro-ph/0302207.








\end{thebibliography}
\end{document}